\newif\iffigs\figstrue
\documentclass[a4paper,11pt]{article}
\usepackage{latexsym,amssymb,lscape,graphics}
\usepackage{graphicx}        
\usepackage{longtable}
\usepackage{multirow}
\usepackage{color}
\usepackage{slashed,epsfig}
\usepackage{amsfonts}

\newtheorem{definizione}{Definition}[section]

\newcommand{\bd}{\begin{definizione}}
\newcommand{\ed}{\end{definizione}}

\def\IC{\relax\,\hbox{$\inbar\kern-.3em{\rm C}$}}
\def\IG{\relax\,\hbox{$\inbar\kern-.3em{\rm G}$}}
\def\IB{\relax{\rm I\kern-.18em B}}
\def\ID{\relax{\rm I\kern-.18em D}}
\def\IL{\relax{\rm I\kern-.18em L}}
\def\IF{\relax{\rm I\kern-.18em F}}
\def\IH{\relax{\rm I\kern-.18em H}}
\def\II{\relax{\rm I\kern-.17em I}}
\def\IN{\relax{\rm I\kern-.18em N}}
\def\IP{\relax{\rm I\kern-.18em P}}
\def\IQ{\relax\,\hbox{$\inbar\kern-.3em{\rm Q}$}}
\def\bfzero{\relax\,\hbox{$\inbar\kern-.3em{\rm 0}$}}
\def\IK{\relax{\rm I\kern-.18em K}}
\def\IG{\relax\,\hbox{$\inbar\kern-.3em{\rm G}$}}
 \font\cmss=cmss10 \font\cmsss=cmss10 at 7pt
\def\IR{\relax{\rm I\kern-.18em R}}
\def\ZZ{\relax\ifmmode\mathchoice
{\hbox{\cmss Z\kern-.4em Z}}{\hbox{\cmss Z\kern-.4em Z}}
{\lower.9pt\hbox{\cmsss Z\kern-.4em Z}} {\lower1.2pt\hbox{\cmsss
Z\kern-.4em Z}}\else{\cmss Z\kern-.4em Z}\fi}
\def\bfone{\relax{\rm 1\kern-.35em 1}}

\def\inbar{\vrule height1.5ex width.4pt depth0pt}
\def\bfzero{\relax{\rm I\kern-.18em 0}}
\def\bfone{\relax{\rm 1\kern-.35em 1}}

\DeclareFontFamily{U}{rsf}{} \DeclareFontShape{U}{rsf}{m}{n}{
  <5> <6> rsfs5 <7> <8> <9> rsfs7 <10-> rsfs10}{}
\DeclareMathAlphabet\Scr{U}{rsf}{m}{n}

\newcommand{\ft}[2]{{\textstyle\frac{#1}{#2}}}
\def\tilde{\widetilde}

\def\1bar{1\hskip -.275cm -}
\def\2bar{2\hskip -.275cm -}
\def\3bar{3\hskip -.275cm -}

\newsavebox{\uuunit}
\sbox{\uuunit}
                 {\setlength{\unitlength}{0.825em}
                      \begin{picture}(0.6,0.7)
                                      \thinlines
                                      \put(0,0){\line(1,0){0.5}}
                                      \put(0.15,0){\line(0,1){0.7}}
                                      \put(0.35,0){\line(0,1){0.8}}
                                     \multiput(0.3,0.8)(-0.04,-0.02){10}{\rule{0.5pt}{0.5pt}}
                      \end {picture}}

\makeatletter \@addtoreset{equation}{section} \makeatother

\def\bfone{\relax{\rm 1\kern-.35em 1}}

\def\bfone{\relax{\rm 1\kern-.35em 1}}
\font\cmss=cmss10 \font\cmsss=cmss10 at 7pt

\newcommand{\su}{\mathfrak{su}}

\begin{document}
\begin{titlepage}
\vskip 0.2cm
\begin{center}
{\Large {\bf  Inflation  and Integrable one-field Cosmologies\\[0.5 cm]
embedded in Rheonomic Supergravity}}\\[1cm]
{ \large Pietro Fr\'e$^{a}$\footnote{Prof. Fr\'e is presently fulfilling the duties of Scientific Counselor of the Italian Embassy in the Russian Federation, Denezhnij pereulok, 5, 121002 Moscow, Russia.}, Alexander S. Sorin$^{b}$ }
{}~\\
{}~\\
\quad \\
{{\em $^{a}$  Dipartimento di Fisica, Universit\'a di Torino,}}
\\
{{\em $\&$ INFN - Sezione di Torino}}\\
{\em via P. Giuria 1, I-10125 Torino, Italy}~\quad\\
{\tt fre@to.infn.it}
{}~\\
\quad \\
{{\em $^{b}$\sl\small Bogoliubov Laboratory of Theoretical Physics}}\\
{{\em  {\tt and} Veksler and Baldin Laboratory of High Energy Physics,}}\\
{{\em Joint Institute for Nuclear Research,}}\\
{\em 141980 Dubna, Moscow Region, Russia}~\quad\\
\emph{e-mail:}\quad {\small {\tt sorin@theor.jinr.ru}}
\quad \\
\end{center}
~{}
\begin{abstract}
In this paper we show that the new approach to the embedding of the inflationary potentials into supergravity, presented in a quite recent paper \cite{minimalsergioKLP} of Ferrara, Kallosh, Linde and Porrati can be formulated within the framework of standard matter coupled supergravity, without the use of the new minimal auxiliary set and of conformal compensators. The only condition is the existence of a translational Peccei Quinn isometry of the scalar K\"ahler manifold. We suggest that this embedding strategy based on a nilpotent gauging amounts to a profound Copernican Revolution. The properties of the inflaton potential are encoded in the geometry of some homogeneous one-dimensional K\"ahler manifolds that now should  be regarded  as the primary object, possibly providing a link with microscopic physics. We present  a simple and elegant formula  for the curvature of the K\"ahler manifold in terms of the potential. Most relevant consequence of the new strategy is that all the integrable potentials quite recently classified in a paper \cite{noicosmoitegr} that we have coauthored, are automatically embedded into supergravity and their associated K\@ahler manifolds demand urgent study. In particular one integrable potential that provides the best fit to PLANCK data seems to have inspiring geometrical properties deserving further study.
\end{abstract}
\end{titlepage}
\tableofcontents
\newpage
\section{Introduction}

Great attention has been recently given to the theoretical interpretation of the results on the CMB spectrum obtained by the European mission PLANCK \cite{Ade:2013uln,Ade:2013zuv}, confirming and extending those of WMAP \cite{Hinshaw:2012aka}.
\par
These analyses have generated a general consensus in the community that the favorite explanation of the observed data is in terms of inflationiary models  based on a single scalar field $\phi$. A vast activity of investigations has produced a series of potentials $V(\phi)$ that display phenomenologically attractive features \cite{Starobinsky:1980te,sasakimukhanov}, can be reconciled with the observed parameters of the CMB power spectrum and have some relation with supergravity \cite{KLpapers}.
\par
It is on the other hand a well known fact that exact analytic solutions of the Friedman equations for the scale-factor $a(t)$ and for a single scalar field $\phi(t)$ with a potential $V(\phi)$ are extremely scarce in the literature.
For that reason, about one year ago, a research programme was started, whose results have been published recently \cite{noicosmoitegr}, aiming at the compilation of a \textit{bestiary of integrable potentials}, namely of a list of one field functions $V(\phi)$ that, once inserted in the Friedman equations, or in their generalization to different gauges, produce an integrable two-field model.
\par
It turned out that the sought for bestiary has a certain amplitude since it includes a few one-parameter series and some sporadic cases. It also displays a challenging general feature: the integrable potentials $V(\phi)$ are always linear combinations, or fractional linear combinations, of a few different exponential functions $\exp[\beta_i \phi]$ with various $\beta_i$. It was immediately recognized that potentials of this type arise naturally in $\mathcal{N}=1$, or extended, supergravity based on target manifolds that are coset spaces $\mathrm{G/H}$ when one gauges them. The scalar potential produced by the gauging  is a polynomial function of the coset representative so that it appears to be of the type described above, whenever it is truncated (possibly in a consistent way) to the Cartan fields.
\par
>From a phenomenological point of view the class of scalar potentials that hosts the integrable ones is also endowed with attractive features. The phenomenon of climbing scalars, which is a generic attribute of this class, was identified in \cite{Dudas:2012vv,Sagnotti:2013ica} as a possible mechanism to explain the low $\ell$ oscillations of the CMB spectrum.
\par
For these reasons, in a parallel way to the compilation of the integrable bestiary, we started an investigation aiming at embedding the integrable potentials in gauged supergravity models. The first results of such a scan, that are going to appear in \cite{nointegrable2}, revealed that such an embedding is a quite hard but not an impossible task since we were able to identify two instances where a suitable truncation of a multifield potential originating from the gauging produces one of our integrable cases. Our results tend to suggest that  integrable potentials are probably excluded in extended supergravity. For instance we were able to classify exhaustively all the gaugings of the $\mathcal{N}=2$, $STU$-model and exclude, within that environment, any truncation to a single scalar field with an integrable potential. We succeeded instead to embed a couple of integrable potentials in $\mathcal{N}=1$ supergravity with a superpotential $W(z)$ of the type produced in Flux Compactifications. Obviously such an embedding is rewarding, since the superpotential that produces the result has in principle an interpretation in terms of Fluxes, opening a way that might eventually link the observed cosmological data to microscopic fundamental physics. Yet, from another  point of view the so far obtained results are not satisfactory for two reasons. Firstly the cases that were successfully embedded into supergravity through a suitable superpotential are not, within the integrable bestiary, cases with particularly attractive features from the phenomenological point of view. Secondly it remains the question why the inflaton driving the history of the Universe should be identified by just that one truncation that is integrable  among the several other available ones that are equally good and equally consistent. In other words the selection of a single inflaton immersed in a multi field model raises problems of naturality.
\par
The authors of a paper recently appeared in the literature \cite{minimalsergioKLP}, have  put forward a new viewpoint that we do not hesitate to describe as a Copernican revolution. Indeed the basic implication of the constructions presented in \cite{minimalsergioKLP} is that we can solve the naturality problem a priori, by  identifying the \textit{inflaton} with the unique scalar field whose potential is generated by a different gauge mechanism, taking advantage of the D-sector, rather than of the F-sector of the Lagrangian, the latter sector being  governed by the superpotential $W(z)$. Further conceptual implications of this Copernican revolution is that the efforts to explain the microscopic origin of phenomenologically good potentials have to shift gear and instead of concentrating on Flux superpotentials they should rather  focus on the properties and the origin of new peculiar and quite intriguing one-dimensional K\"ahler manifolds that happen to replace the familiar upper complex plane, endowed with the time honored Poincar\`e Lobachevsky metric. Furthermore,  implementing this revolution, all the positive definite integrable potentials included in the bestiary of \cite{noicosmoitegr} become immediately available to supergravity and actually are put into one-to-one correspondence with the non homogeneous one-dimensional K\"ahler manifolds alluded to, few lines above.
\par
Although what we described in the previous paragraph  is what is actually at the stake in the results of \cite{minimalsergioKLP}, the emphasis in \cite{minimalsergioKLP} and in a few previous papers \cite{Ketov:2010qz,Ketov:2012jt,johndimitri,Kallosh:2013lkr,Kallosh:2013hoa,Farakos:2013cqa}, that prepared the road for this
big leap, is somehow shifted to another issue that we consider unessential and we fear might distract the attention of the reader putting him on a  sterile conceptual track. As it is evident also from the title of \cite{minimalsergioKLP} (\textit{Minimal Supergravity Models of Inflation}) it seems to be advocated by its authors that the key ingredient of supergravity that introduces  the new stand point  allowing to embed almost arbitrary one-field potentials is the use of the New Minimal Formulation of the off-shell theory. This view is further sustained by the very means of construction of the presented models that relies on superfields, conformal tensor calculus and the use of a linear multiplet compensator rather then a chiral multiplet one. One of the goals of our paper is to show that all that is unnecessary and that the new stand point can be realized within the framework of standard matter coupled supergravity \cite{standardN1} with just second derivatives. The only crucial requirement is the existence of a Peccei-Quinn like translation isometry in the K\"ahler manifold encompassing all the scalars,  the inflaton being included in this number. The gauging of this translational symmetry selects the inflaton and separates it from his friends.
 \par
 In order to better clarify the issue of the auxiliary fields involved in the construction, we utilize the rheonomic approach to supergravity \cite{castadauriafre2} and we recall the results of paper \cite{D'Auria:1988qm} that appeared twenty-five  years ago, shortly after the papers \cite{Cecotti:1987sa,Cecotti:1987qe} that form the basis for the model construction presented in \cite{minimalsergioKLP}. According to the rheonomic conception where the auxiliary fields are simply black-boxes parameterizing the superfield curvatures, that eventually,  become expressed in terms of the physical fields while going on-shell, in \cite{D'Auria:1988qm} it was shown that the New Minimal and the Old Minimal formulation coexist within a larger $16\oplus 16$ parameterization that is actually utilized in matter coupled supergravity to insert all the fields and all the couplings. The New Minimal formulation corresponds to the possibility of eliminating the complex scalar auxiliary field $S$ by means of a Weyl transformation. The condition for the existence of such a Weyl transformation is the presence, for the K\"ahler manifold of the scalars, of a Peccei Quinn translation isometry. Indeed the appropriate parameter of the Weyl transformation mentioned above is precisely the prepotential of the Killing vector that generates  the Peccei-Quinn translations. Significantly the condition of the Peccei Quinn symmetry is the same condition that allows to embed general one-field potentials via a gauging procedure. When it is there, performing the Weyl transformation to the new minimal form of the superspace curvature is unessential (actually it takes us away from the natural Einstein frame): important is just the presence of such a symmetry.
 \par
 Having clarified this in the second part of the paper we dwell on the consequences of the Copernican revolution. The gauging procedure puts into correspondence every choice of a positive definite potential with the choice of a different K\"ahler geometry. In particular we derive a simple but inspiring relation between the potential and the curvature of the corresponding K\"ahler manifold. Applied to some of the phenomenologically most promising integrable potentials of our bestiary, this formula reveals that the corresponding curvature realizes a smooth transition between two asymptotic  Lobachevsky planes with different curvatures, respectively approached at large and small coordinates. In between there is a kink or wiggle structure which is probably responsible for most of the physical consequences on inflation.
 \par
 The alert we would like to put forward with the present paper is that an understanding of the Fundamental Theory behind the scene hidden by each more or less successful inflationary potential necessarily passes through a comprehension of the geometric structure and origin of these new strange K\"ahler manifolds.
\section{Matter coupled $\mathcal{N}=1$ Supergravity in the rheonomy framework}
In order to discuss the main points announced in the introduction, in the present section we summarize the formulation of $\mathcal{N}=1$, $D=4$ supergravity  and its coupling to matter multiplets according to rheonomy approach. (For a general presentation of this approach that dates back to the beginning of the eighties, see the book \cite{castadauriafre2}). Since the structure of the Lagrangian is codified into the geometric structures associated with the K\"ahler geometry of the manifold $\mathcal{M}_K$ that contains the $n$ complex scalar fields spanning the bosonic sector of the Wess-Zumino multiplets, we begin by recalling the basic geometric ingredients utilized in the construction of the theory. This is done for the reader's convenience and also in order to fix our conventions.
\subsection{K\"ahler Geometry and Hodge bundle ingredients}
The local  geometry of a  K\"ahler manifold $\mathcal{M}_K$ is encoded in the K\"ahler potential $\mathcal{K}(z,z)$ that is a real function of the complex coordinates $z^i$ and $z^{j^\star}$. The hermitian K\"ahler metric is given by:
\begin{equation}\label{kelerusmetrus}
    g_{ij^\star} \, = \, \partial_i\partial_{j^\star} \, \mathcal{K} \quad \Rightarrow \quad ds^2_{K} \, = \, \partial_i\partial_{j^\star}\mathcal{ K} \, dz^i \otimes dz^{j^\star}
\end{equation}
and the Levi Civita connection on the K\"ahler manifold as the following form:
\begin{eqnarray}\label{levicivita}
    \Gamma^i_{\phantom{k}k} & = & \left\{\begin{array}{c}
                                             i \\
                                             j \,k
                                           \end{array}\right\}\,dz^j \, = \, g^{i\ell^\star} \, \partial_j \, g_{\ell^\star k} \, dz^j \nonumber\\
    \Gamma^{i^\star}_{\phantom{k}k^\star} & = & \mbox{Complex Conjugate of } \, \Gamma^i_{\phantom{k}k}
\end{eqnarray}
The manifold is Hodge-K\"ahler if there exists a line bundle $\mathcal{L} \rightarrow \mathcal{M}_K$ whose Chern class coincides with the K\"ahler class, namely with the cohomology class of the K\"ahler two-form:
\begin{equation}\label{callaclassa}
    \mathrm{K} \, = \, {\rm i} \, g_{ij^\star} \,dz^i \, \wedge \, dz^{j^\star}
\end{equation}
Explicitly we must have $c_1(L) \, = \,\left[\mathrm{K}\right]$, where the bracket denotes the cohomology class of the closed $p$-form embraced by it. The holomorphic sections of this line bundle are the possible superpotentials that encode the self interactions of the Wess-Zumino multiplets and their coupling to supergravity. The exponential of the K\"ahler potential is a fiber metric on the Hodge bundle:  for any holomorphic section $W(z)$ of such a bundle we define an invariant  norm by means of the following position:
\begin{equation}\label{sectionnorm}
    || W||^2 \, = \, \exp[\mathcal{K}] \, W(z) \, \overline{W}(\bar{z})
\end{equation}
A fundamental object entering the construction of matter coupled supergravity is the logarithm of the superpotential norm:
\begin{equation}\label{zittozitto}
    G(z,\bar{z}) \, = \, \log \, || W||^2 \, = \, \mathcal{K} \, + \, \log \, W \, + \, \log \overline{W}
\end{equation}
Another fundamental ingredient in the matter coupling construction and in its gauging is provided by the prepotentials of the holomorphic Killing vectors. Following the discussion and the conventions of \cite{NoistandardN2}, if $k^i_\Lambda (z)$, together with its complex conjugate $k^{i^\star}_\Lambda (\bar{z})$, are holomorphic Killing vectors, in the sense that the transformation:
\begin{equation}\label{transformazione}
    z^i \, \rightarrow \, z^i \, + \, \epsilon^\Lambda \, k^i_\Lambda (z)
\end{equation}
is an infinitesimal isometry of the K\"ahler metric for all choices of the small parameters $\epsilon^\Lambda$, then the prepotentials of these Killing vectors, which realize the corresponding isometry Lie algebra as a Lie-Poisson algebra on the K\"ahler manifold, are the real functions $\mathcal{P}_\Lambda(z,\bar{z})= \mathcal{P}_\Lambda(z,\bar{z})^\star$ defined by the following relations:
\begin{equation}\label{gospadi}
    k^i_\Lambda (z) \, = \, {\rm i} \, g^{ij^\star} \, \partial_{j^\star} \, \mathcal{P}_\Lambda \quad ; \quad k^{j^\star}_\Lambda (\bar{z}) \, = \, - \, {\rm i}
     \, g^{ij^\star} \, \partial_i \, \mathcal{P}_\Lambda
\end{equation}
In terms of the $G$ function, supposedly invariant under the considered isometries, the Killing vector prepotentials, satisfying the defining condition (\ref{gospadi}), are constructed through the following formula:
\begin{equation}\label{bozhemoi}
    \mathcal{P}_\Lambda \, = \, -\,{\rm i} \, \ft 12 \, \left(k^i_\Lambda \, \partial_i \,G \,  - \, k^{i^\star}_\Lambda \,\partial_{i^\star} \,G \right)
\end{equation}
\subsection{Sections of the Hodge bundle  and the fermions}
The basic geometric mechanism that allows to gauge the global symmetries of supergravity coupled to $n$ Wess Zumino multiplets is the so named gauging of the composite connections. Let us recall such a notion.
The isometries of the K\"ahler metric  that take the infinitesimal form (\ref{transformazione}) extend to global symmetries of the full theory, including also the fermions, since all the items appearing in the lagrangian transform covariantly. From the geometrical point of view all fields are sections of the tangent bundle to the K\"ahler manifold and at the same time they are also sections of appropriate powers of the Hodge bundle. The subtle point is that under a holomorphic isometry:
\begin{equation}\label{giacomino}
    z^i \, \rightarrow\, \hat{z}^i \, = \,f^i(z)
\end{equation}
the K\"ahler potential does not necessarily remain invariant rather it transform as follows:
\begin{equation}\label{ginocchio}
    \mathcal{K}\left(\hat{z},\hat{\bar{z}}\right) \, = \, \mathcal{K}\left({z},{\bar{z}}\right) \, + \, F(z) \, + \bar{F}(\bar{z})
\end{equation}
where $F(z)$ is some holomorphic function associated with the considered transformation. The function $G$ and the norm  (\ref{sectionnorm}) are effectively invariant under isometries if the superpotential $W(z)$ transform as follows:
\begin{equation}\label{fontanafredda}
    W(\hat{z}) \, = \, W({z}) \, \exp[- F(z)]
\end{equation}
By definition, the above transformation property is what defines a section of the Hodge line-bundle of weight $p=1$. The fermion fields, namely the gravitino $\psi$, the chiralinos $\chi^i$, $\chi^{j^\star}$ and the gauginos
$\lambda^\Lambda$ transform also as sections of the Hodge bundle, yet with half integer weights that we presently spell off.
\par
According to \cite{castadauriafre2}, we introduce the following notation for the chiral projections of the gravitino one-form $\psi$ and of the gaugino 0-forms $\lambda^\Lambda$ that are Majorana:
\begin{eqnarray}\label{gorgera}
    \psi & = & \psi_\bullet \, + \, \psi^\bullet\quad ; \quad \left. \{ \begin{array}{rcl}
                                                                          \gamma_5 \, \psi_\bullet & =&\psi_\bullet \\
                                                                          \gamma_5 \, \psi^\bullet & =&-\, \psi^\bullet
                                                                        \end{array}
    \right.\nonumber\\
    \lambda^\Lambda & = & \lambda^\Lambda_\bullet \, + \, \lambda^{\Lambda|\bullet}\quad ; \quad \left.\{ \begin{array}{rcl}
                                                                          \gamma_5 \, \lambda^\Lambda_\bullet & =&\lambda^\Lambda_\bullet \\
                                                                          \gamma_5 \, \lambda^{\Lambda|\bullet} & =&-\, \lambda^{\Lambda|\bullet}
                                                                        \end{array}\right.
\end{eqnarray}
while for the complex chiralinos we simply have:
\begin{equation}\label{fermentilattici}
    \gamma_5 \, \chi^i \, = \, \chi^i \quad ; \quad \gamma_5 \, \chi^{j^\star} \, = \, - \,\chi^{j^\star}
\end{equation}
Having clarified the notation the appropriate Hodge transformations for the fermions are:
\begin{equation}\label{pasticcaleone}
\begin{array}{ccccccc}
       \psi_\bullet & \to & \exp\left[{\rm i}\, \ft 12 \, F(z) \right]\, \psi_\bullet &\quad ; \quad&\psi^\bullet & \to & \exp\left[-\, {\rm i}\, \ft 12 \,  F(z) \right]\,  \psi^\bullet \\
       \lambda^\Lambda_\bullet & \to & \exp\left[{\rm i} \, \ft 12 \,  F(z) \right] \,\lambda^\Lambda_\bullet  &\quad ; \quad & \lambda^{\Lambda|\bullet}& \to & \exp\left[-\, {\rm i}\, \ft 12 \,  F(z) \right] \, \lambda^{\Lambda|\bullet} \\
       \chi^i & \to &\exp\left[- \, {\rm i}\, \ft 12 \,  F(z) \right] \, \chi^i  & ; & \chi^{j^\star} & \to  & \exp\left[{\rm i}\, \ft 12 \,  F(z) \right] \,\chi^{j^\star}
     \end{array}
\end{equation}
These transformations are compensated by the transformation of the Hodge bundle connection which is the following composite one-form:
\begin{equation}\label{HodgeKalloconno}
    Q \, \equiv \, {\rm i} \, \ft 12 \, \left( \partial_i \mathcal{K}\, dz^i \, - \, \partial_{j^\star} \mathcal{K} \, dz^{j^\star}\right)
\end{equation}
and enters the covariant derivatives of the fermions. For instance the gravitino covariant derivative is defined as follows:
\begin{equation}\label{giannibeffa}
    \nabla \, \psi_\bullet \, = \, \mathcal{D}\psi_\bullet  \, + \, {\rm i} \, Q \, \wedge \, \psi \quad ; \quad \mathcal{D}\psi_\bullet \, = \, d\psi \, - \, \ft 14 \, \omega^{ab} \, \wedge \, \gamma_{ab} \, \psi_\bullet
\end{equation}
The gravitino one-form and the gaugino zero-forms have no indices along the tangent bundle of the K\"ahler manifold and therefore do not transform in the canonical bundle. On the other hand the chiralinos carry tangent space indices and with respect to the canonical bundle transform  as holomorphic vectors. Correspondingly they enter the lagrangian covered by a covariant derivative of the form:
\begin{equation}
\nabla \,\chi^i \, \equiv\,  \mathcal{D}\chi^i \, + \, {\Gamma}^i_{\phantom{i}k} \chi^j
\,  - \, {\rm i} \ft 12 \, {Q} \, \chi^i \label{dchipredefi}
\end{equation}
In this way the isometries of the K\"ahler manifold are promoted to global symmetries of supergravity coupled to $n$ vector multiplets.
\subsection{Gauging of the composite connections}The basic geometric mechanism that allows to gauge the above described global symmetries is the so named \textit{gauging of the composite connections}. Let us recall such a notion, according to the discussion of \cite{NoistandardN2}. There the construction was applied to $\mathcal{N}=2$ supergravity so that the composite connections to be gauged were those emerging in Special K\"ahler Geometry. Here we focus on $\mathcal{N}=1$ supergravity and we just have Hodge-K\"ahler manifolds, yet the procedure is completely identical and it was already introduced in \cite{castadauriafre2}, but only for symmetries that are linearly realized on the scalars. Here we smootly generalize it to any type of holomorphic isometry, by means of the prepotential of the Killing vectors. The connections to be gauged are two: the Hodge-K\"ahler connection
(\ref{HodgeKalloconno})  and the Levi-Civita connection (\ref{levicivita}). We set :
\begin{eqnarray}
                Q &\to & \widehat{Q} \, \equiv \, {\rm i} \, \ft 12 \, \left( \partial_i \mathcal{K}\, \nabla z^i \, - \, \partial_{j^\star} \mathcal{K} \, \nabla z^{j^\star}\right)\\
                \Gamma^i_{\phantom{i}j} &\to & \widehat{\Gamma}^i_{\phantom{i}j}  \, = \, \left\{\begin{array}{c}
                                             i \\
                                             k \,j
                                           \end{array}\right\}\,\nabla z^k
              \end{eqnarray}
where
\begin{equation}\label{cirimella}
    \nabla z^k  \, = \, dz^k \, + \, g\, \mathcal{A}^\Lambda \, k^i_\Lambda(z)
\end{equation}
It follows from the various identities presented above that:
\begin{eqnarray}
  \widehat{Q}&=& Q \, + \, g \, \mathcal{A}^\Lambda \, \mathcal{P}_\Lambda\\
  \widehat{\Gamma}^i_{\phantom{i}j}  &=& {\Gamma}^i_{\phantom{i}j} \, + \,\mathcal{A}^\Lambda \, \partial_{j}\,k^i_\Lambda(z) \label{bertoldino}
\end{eqnarray}
\subsection{The $\mathcal{N}=1$ Supergravity curvatures and their rheonomic off-shell parameterizations}
Having prepared the stage with the previously listed geometric ingredients, the next step in the rheonomic algorithm for the construction of supergravity theories is given by the definition of the supergravity curvatures and by the rheonomic solution of the associated Bianchi identities.
\par
In the $\mathcal{N}=1$ case the curvatures of the superPoincar\'e algebra, extended by means of a $\mathrm{U(1)}$ R-symmetry acting only the fermionic charge and, correspondingly, on its dual one-form, the gravitino $\psi$, are the following ones:
\begin{eqnarray}
  \mathfrak{R}^a &\equiv& \mathcal{D} \, V^a \, - \, {\rm i} \bar{\psi}^\bullet \, \wedge \,\gamma^a \, \psi_\bullet \quad; \quad
\left(  \mathcal{D} \, V^a  \,\equiv \, dV^a \, - \, \omega^{ab} \, \wedge \, V_b \right ) \label{torsiondefi} \\
  \mathfrak{R}^{ab}&\equiv& d\omega^{ab} \, - \, \omega^{ac} \, \wedge \, \omega^{cb}\label{riemandefi} \\
  \rho_\bullet &\equiv& \mathcal{D} \, \psi_\bullet \, + \, {\rm i} \ft 12 \, \mathbf{A} \, \wedge \, \psi_\bullet \quad ; \quad  \left (
  \mathcal{D} \, \psi_\bullet   \,\equiv \, d\psi_\bullet \, - \, \ft 14 \, \gamma_{ab} \, \omega^{ab} \, \wedge \, \psi_\bullet \right) \label{rhodefidown}\\
   \rho^\bullet &\equiv& \mathcal{D} \, \psi^\bullet \, + \, {\rm i} \ft 12 \, \mathbf{A} \, \wedge \, \psi^\bullet \quad ; \quad  \left (
  \mathcal{D} \, \psi^\bullet   \,\equiv \, d\psi^\bullet \, - \, \ft 14 \, \gamma_{ab} \, \omega^{ab} \, \wedge \, \psi^\bullet \right) \label{rhodefiup}\\
  \mathfrak{R}[\mathbf{A}] &\equiv& d \mathbf{A} \label{RdiAdefi}
\end{eqnarray}
Setting these curvature to zero one obtains the Maurer Cartan equations that constitute the dual description of the super Poincar\'e Lie algebra. At non vanishing curvatures the members of the one-form connection, acquire the physical interpretation of  fields associated with the graviton multiplet. Specifically $V^a$ is the vielbein, $\psi$ the gravitino, $\omega^{ab}$ the spin connection and $\mathbf{A}$ is an auxiliary one form that will be identified with the composite Hodge-K\"ahler connection in the coupling to matter.
The Bianchi identities following from eq.s (\ref{torsiondefi}-\ref{RdiAdefi}) are the following ones:
\begin{eqnarray}
 \mathcal{D} \mathfrak{R}^a \, + \, \mathfrak{R}^{ab} \, \wedge \, V_b \, - \, {\rm i} \left( \bar{\psi}^\bullet \, \wedge \, \gamma^a \, \rho_\bullet
 \, + \, \bar{\psi}_\bullet \, \wedge \, \gamma^a \, \rho^\bullet \right) & = &0 \label{torsionBianchi} \\
  \mathcal{D} \mathfrak{R}^{ab}&= & 0 \label{riemanBianchi} \\
 \mathcal{D} \rho_\bullet \, + \, \ft 14 \, \gamma_{ab} \, \mathfrak{R}^{ab} \,\gamma_{ab} \, \wedge \, \psi_\bullet \, - \, {\rm i} \, \ft 12 \mathfrak{R}[A] \, \wedge \psi_\bullet & = & 0 \label{rhoBianchidown}\\
  \mathcal{D} \rho^\bullet \, + \, \ft 14 \, \gamma_{ab} \, \mathfrak{R}^{ab} \,\gamma_{ab} \, \wedge \, \psi^\bullet \, - \, {\rm i} \, \ft 12 \mathfrak{R}[A] \, \wedge \psi^\bullet & = & 0 \label{rhoBianchiup}\\
  d \,\mathfrak{R}[\mathbf{A}] &=& 0 \label{RdiABianchi}
\end{eqnarray}
As strongly emphasized in \cite{D'Auria:1988qm} and then widely ridiscussed in the book \cite{castadauriafre2}, the peculiarity of the $\mathcal{N}=1$ case is that the rheonomy principle, which demands that the outer components of the superspace curvatures (eq.s (\ref{torsiondefi}-\ref{RdiAdefi}) in the present case) be expressed in terms of inner components\footnote{In recent literature on the conjectured finiteness of supergravity theories the time honored rheonomy approach has been somewhat rediscovered and what we named for almost thirty years the inner and outer components of the superspace curvatures  are now fashionably and quite imaginatively renamed the \textit{body} and the \textit{soul}.} or other space-time fields, can be reconciled with the Bianchi-identities in an off-shell way, by introducing a finite list of black-box tensor structures in terms of which the Bianchi identities (\ref{torsionrheo}-\ref{RdiArheo}) can be exactly solved without implying any further constraint on the Riemann tensor or the gravitino field strength. This is what people call an off-shell formulation of supergravity and the black-box tensor structures that allow for the solution of Bianchi.s correspond in other approaches to the notion of auxiliary fields. In the early paper \cite{D'Auria:1988qm} it was stressed that a convenient non minimal solution of the Bianchi identities that allows for the embedding of both the so named \textbf{old-minimal} and the \textbf{new minimal} set of auxiliary fields is provided in terms of $16\oplus 16$ off-shell degrees of freedom.
The $16\oplus 16$ rheonomic parameterization of the curvatures is the following one:
\begin{eqnarray}
\mathfrak{R}^a &=& 0\label{torsionrheo} \\
  \mathfrak{R}^{ab}& =& R^{ab}_{\phantom{ab}mn} \, V^m \, \wedge \, V^n \, - \, \bar{\psi}^\bullet \theta_\bullet^{ab|c} \, \wedge \, V_c \, - \, \bar{\psi}_\bullet \theta_c^{ab|\bullet} \, \wedge \, V^c \nonumber\\
 && \, - \, {\rm i} \, S^\star \, \bar{\psi}_\bullet \, \wedge \, \gamma^{ab} \, \psi_\bullet \, + \, {\rm i} \, S\, \bar{\psi}^\bullet \, \wedge \, \gamma^{ab} \, \psi^\bullet \nonumber\\
  && \, - \, 2 \,{\rm i} \,A^\prime_c \, \bar{\psi}^\bullet \, \gamma_d \, \psi_\bullet \, \epsilon^{abcd} \label{riemanrheo} \\
  \rho_\bullet &=& \rho_\bullet^{ab} \, V_a \, \wedge \, V_b \, + \, {\rm i} \, A_a \, \psi_\bullet \, \wedge \, V^a \, + \, {\rm i} \, A_a^\prime \, \gamma^{ab} \, \psi_\bullet \, \wedge \, V^b \nonumber\\
  && \, - \, \psi_\bullet \, \wedge \, \bar{\psi}^\bullet \, \zeta^\bullet
\label{rhorheodown}\\
   \rho_\bullet &=& \rho^\bullet_{ab} \, V^a \, \wedge \, V^b \, + \, {\rm i} \, A_a \, \psi^\bullet \, \wedge \, V^a \, + \, {\rm i} \, A_a^\prime \, \gamma^{ab} \, \psi^\bullet \, \wedge \, V^b \nonumber\\
  && \, - \, \psi^\bullet \, \wedge \, \bar{\psi}_\bullet \, \zeta_\bullet
\label{rhorheoup}\\
  \mathfrak{R}[\mathbf{A}] &=& K_{ab} \, V^a \, \wedge \, V^b \, - \, \bar{\psi}_\bullet \, \Phi_{a|\bullet} \, \wedge \, V^a
  \, - \, \bar{\psi}^\bullet \, \Phi_a^\bullet \, \wedge \, V^a \label{RdiArheo}
\end{eqnarray}
In the above formulae the black boxes that constitute the auxiliary fields are:
\begin{enumerate}
  \item The complex scalar $S$ encoding 2 bosonic degrees of freedom
  \item The Lorentz real vector  $A_a^\prime$ encoding $4$ bosonic degrees of freedom.
  \item The gauge one form $\mathbf{A}$ encoding $3$ bosonic degrees of freedom
  \item The spinor $\zeta$ encoding $4$ fermionic degrees of freedom.
  \item A new scalar $C$ appearing in the parameterization of the derivative of $\zeta$:
  \begin{equation}\label{guliardo}
    \nabla \zeta \, = \, \nabla_a\zeta \, V^a \, + \, C \, \psi
  \end{equation}
  which provides $1$ bosonic degree of freedom.
  \end{enumerate}
  It is important to note that the vector $A_a$ does not constitute an independent degree of freedom because it can always be reabsorbed into a redefinition of the gauge form $\mathbf{A}$ by setting:
  \begin{equation}\label{ciaccio}
    \mathbf{A}^\prime \, = \, \mathbf{A} \, + \, A_a \, V^a
  \end{equation}
  and so doing it disappears from the rheonomic parameterizations.
\par
If we add the $10$ bosonic auxiliary degrees of freedom to the $6$ off-shell degrees of freedom of the graviton we obtain $16$ bosons. We recall that $6=10-4$ where $10$ are the entries of a symmetric tensor and $4$ are the degrees of freedom that can be removed by gauge transformations which are diffeomorphisms in this case.
Similarly, if we add the $4$ fermionic auxiliary degrees of freedom to the $12$ off-shell degrees of freedom of the gravitino we obtain $16$ fermions.  We remind that $12=16 - 4$ where $16$ are the entries of a spinor-vector and $4$ are the degrees of freedom that can be removed by local supersymmetry.
\par
The spinor-tensor structures $\theta_c^{ab|\bullet} $, $\theta_\bullet^{ab|c}$, $\Phi_a^\bullet$, $\Phi_{a|\bullet}$ are not independent objects, rather they are linear combinations of the gravitino field strength $\rho_{ab}$ and of the auxiliary $\zeta$. Their explicit form was exactly calculated in \cite{D'Auria:1988qm} and we refer the interested reader to that paper, since we do not need them for our present goals.
\par
As explained in \cite{D'Auria:1988qm}, the Old-minimal and the New-minimal off-shell formulations of supergravity correspond to two different  truncations of the non minimal $16\oplus16$ solution of Bianchi identities  to a  $12\oplus12$ minimal one that is still consistent and off-shell (no field equations are implied for the graviton or the gravitino). From this point of view there is nothing particularly profound or challenging in the choice of one or the other formulation, as instead it seems to be advocated when the constructive approach to supergravity is based on superfields or superconformal tensor calculus \cite{Cecotti:1987qe},\cite{Cecotti:1987sa}\cite{minimalsergioKLP}.
\begin{description}
  \item[A)] The Old Minimal formulation is obtained by suppressing altogether in the definition of the curvatures and in the parameterization (\ref{torsionrheo}-\ref{RdiArheo}) the one form $\mathbf{A}$ (and hence also  $A_a$),  the spinor  $\zeta$ and the scalar $C$ related to it. In these we leave $S$  and $A_a^\prime$,  which constitute the Old Minimal Set, as unique carriers of the missing bosonic degrees of freedom.
  \item[B)] The New Minimal set is obtained instead by suppressing $S$, $\theta$ and $C$, leaving the gauge form $\mathbf{A}$ and $A_a^\prime$ that is now subject to a  divergenless  constraint:
      \begin{equation}\label{barione}
        \mathcal{D}^a A_a^\prime \, = \, 0
      \end{equation}
      \end{description}
   The constraint (\ref{barione}) which reduces the degrees of  freedom of  $A_a^\prime$  from $4$ to $3$ can be solved by assuming that $A_a^\prime$ is the dual of the field strength of a two-form:
   \begin{equation}\label{soluzionechimica}
    A_a^\prime \, \sim \, \epsilon_{abcd} \, \,H^{bcd}
   \end{equation}
 and this leads to an interesting formulation of the New Minimal model in terms of a Free Differential Algebra  as
 pointed out in \cite{contownsend}.
\par
In \cite{D'Auria:1988qm} it was shown that a rheonomic parameterization of the New Minimal type can be reached starting from a parameterization of the Old Minimal type by performing a suitable field dependent Weyl transformation that cancels the $S$ terms in the parameterization of the curvatures. When the theory is already matter coupled to Wess-Zumino multiplets and the auxiliary fields are expressed in terms of physical fields (as we are going to show below), the existence of the suitable Weyl field-dependent parameter mentioned above is guaranteed by the existence of a Peccei-Quinn translational isometry in the considered K\"ahler manifold. Indeed as shown in \cite{D'Auria:1988qm}, the suitable Weyl parameter is nothing else but the prepotential of the translation symmetry Killing vector. Therefore the real meaning of \textit{New Minimal Formulation} is the existence of such a Peccei-Quinn symmetry. No new result is obtained by actually performing the above mentioned Weyl transformation which makes all the formulae clumsier and less practical.
\par
On the other hand the clue to the embedding of inflationary models into supergravity is critically based on the existence of the quoted Peccei Quinn translational symmetry which is truely the key point.
\par
It follows from the above discussion that in order to implement the Copernican revolution that shifts the inflaton potential from the $F$-sector to the D-sector of the supergravity lagrangian and opens new challenging views on integrable cosmologies, there is no need of a minimal formulation in terms of  superfields or of complicated manipulations in conformal tensor calculus. The standard rheonomic or component formulation of matter coupled $\mathcal{N}=1$ supergravity are perfectly apt to incorporate the essential mechanism and actually better clarify its nature.
\par
So we turn to recall the essential features of matter coupled $\mathcal{N}=1$ supergravity in the rheonomic approach.
\subsection{Matter curvatures of the Wess Zumino multiplets and of the gauge sector}
The third step in the rheonomic construction of a supergravity theory consists of introducing \textit{matter curvatures} or, if one prefers such a language, covariant derivatives of the matter fields which have to be duely given a consistent rheonomic parameterization. In the most general $\mathcal{N}=1$ theory we have $n$ Wess-Zumino multiplets whose scalar fields span some Hodge-Kahler manifold $\mathcal{M}_K$ while their chiralino partners are sections both of the Hodge line bundle and of the canonical tangent bundle, as we have already explained. In addition we have $m$ vector multiplets enumerated by an index $\Lambda \, =\, 1, \dots , m$, each of which comprises a gauge one form $\mathcal{A}^\Lambda$  and a Majorana gaugino $\lambda^\Lambda$.
\par
Henceforth we introduce the following matter curvatures for the WZ multiplets
\begin{eqnarray}
\nabla \,z^i &\equiv& dz^i \, + \, g \, \mathcal{A}^\lambda  \, k^i_\Lambda(z) \label{dzdefi}\\
 \nabla \,z^{j^\star} &\equiv& dz^{j^\star} \, + \, g \, \mathcal{A}^\lambda  \, k^{j^\star}_\Lambda(\bar{z}) \label{dzdefistar}\\
\nabla \,\chi^i &\equiv& \mathcal{D}\chi^i \, + \, \widehat{\Gamma}^i_{\phantom{i}k} \chi^j\,  - \, {\rm i} \ft 12 \, \widehat{Q} \, \chi^i \label{dchidefi}\\
 \nabla \,\chi^i &\equiv& \mathcal{D}\chi^i \, + \, \widehat{\Gamma}^i_{\phantom{i}k} \chi^j\,  - \, {\rm i} \ft 12 \, \widehat{Q} \, \chi^i \label{dchidefistar}
\end{eqnarray}
where $\widehat{Q}$ and $\widehat{\Gamma}^i_{\phantom{i}k}$ are the gauged connections defined in eq.(\ref{bertoldino}) and
\begin{eqnarray}
\mathfrak{F}^\Lambda &\equiv& d\mathcal{A}^\Lambda\, + \, g \, f^{\Lambda}_{\phantom{\Lambda}\Gamma\Delta} \, \mathcal{A}^\Gamma \, \wedge \, \mathcal{A}^\Delta \label{FAdefi}\\
  \nabla \, \lambda^\Lambda_\bullet  & \equiv & \mathcal{D}\,\lambda^\Lambda_\bullet  \, + \, {\rm i} \, \ft 12 \, \widehat{Q} \, \lambda^\Lambda_\bullet
  \label{DLlambdadefiup}\\
  \nabla \, \lambda^{\Lambda|\bullet } & \equiv & \mathcal{D}\,\lambda^{\Lambda|\bullet } \, + \, {\rm i} \, \ft 12 \, \widehat{Q} \, \lambda^{\Lambda|\bullet }
  \label{DLlambdadefidown}
\end{eqnarray}
for the gauge sector, where $f^{\Lambda}_{\phantom{\Lambda}\Gamma\Delta} $ are the structure constants of a completely generic gauge group.
\par
The rheonomic parameterizations of these matter curvatures is the following one:
\begin{eqnarray}
\nabla \,z^i &=& Z_a^{i} \, V^a \, + \, \bar{\psi}_\bullet \, \chi^i \label{dzrheo}\\
 \nabla \,z^{j^\star} &=& \bar{Z}_a^{j^\star} \, V^a \, + \, \bar{\psi}^\bullet \, \chi^{j^\star}\label{dzrheostar}\\
\nabla \,\chi^i &\equiv& \nabla_\chi^i \,V^a \,  +\,{\rm i} Z^i_a \, \gamma^a \, \psi^\bullet \, + \, \mathcal{H}^i \psi_\bullet \label{dchirheo}\\
\nabla \,\chi^{j^\star} &\equiv& \nabla_\chi^{j^\star} \,V^a \,  +\,{\rm i} \bar{Z}^{j^\star}_a \, \gamma^a \, \psi_\bullet \, + \, \mathcal{H}^{j^\star} \psi_\bullet \label{dchirheostar}
\end{eqnarray}
for the WZ sector and
\begin{eqnarray}
\mathfrak{F}^\Lambda &=& F^\Lambda\,_{ab} \, V^a \, \wedge \, V^b \, + \, {\rm i} \bar{\lambda}^\Lambda_\bullet \,
\gamma_m \, \psi^\bullet \, \wedge \, V^m \, +  \, {\rm i} \bar{\lambda}^{\Lambda|\bullet} \,
\gamma_m \, \psi_\bullet \, \wedge \, V^m  \label{FArheo}\\
  \nabla \, \lambda^\Lambda_\bullet  & = & \nabla_a\,\lambda^\Lambda_\bullet  \, V^a  \, + \, F^{+|\Lambda}_{ab} \, \gamma^{ab} \, \psi_\bullet \, + \, {\rm i} \, \mathcal{D}^\Lambda \, \psi_\bullet
  \label{DLlambdarheoup}\\
  \nabla \, \lambda^{\Lambda|\bullet } & = & \nabla_a\,\lambda^{\Lambda|\bullet }  \, V^a  \, + \, F^{-|\Lambda}_{ab} \, \gamma^{ab} \, \psi^\bullet \, + \, {\rm i} \, \mathcal{D}^\Lambda \, \psi^\bullet
  \label{DLlambdarheodown}
\end{eqnarray}
for the gauge sector.
\par
The essential newcomers entering the game when introducing matter are the new auxiliary fields:
\begin{itemize}
  \item the $n$ complex scalars $\mathcal{H}^i$ appearing in the outer direction (soul) parameterization of the chiralino curvatures
  \item the $m$ real scalars $D^\Lambda$ appearing in the outer direction (soul) parameterization of the gaugino curvatures
\end{itemize}
The matter coupling is achieved when all the auxiliary fields entering the supergravity and matter curvatures are expressed in terms of the matter fields in a way consistent with Bianchi identities (up to field equations, since by such an identification we automatically go on shell).
\subsection{On shell form of the Auxiliary Fields}
Following the construction of the $\mathcal{N}=1$ theory explained in the book \cite{castadauriafre2}, which was repeated with more refinements for the $\mathcal{N}=2$ case in \cite{NoistandardN2}, the  supergravity auxiliary fields take the following form:
\begin{eqnarray}
  S &=& {\rm i} \, e \, \exp \, \left [ \ft 12 \, G\right ] \, = \,{\rm i} \, e \,\exp \, \left [ \ft 12 \, \mathcal{K}\right ] \, \sqrt{W\,\overline{W}} \label{Sfildo}\\
  A_a &=& \mu \, \tilde{T}_a \quad ; \quad  \tilde{T}_a \, \equiv \, \mathrm{Im}\mathcal{N}_{\Lambda\Sigma} \, \bar{\lambda}^\Lambda_\bullet \, \gamma_a \lambda^{\Sigma|\bullet}\label{Afildo} \\
  A_a^\prime &=& \ft 18 \, T_a \, + \, \ft 12 \, \nu \, \tilde{T}_a \quad ; \quad T_a  \, = \, g_{ij^\star} \, \bar{\chi}^i \, \gamma_b \chi^{j^\star}\label{Aprimofildo}\\
  \zeta_\bullet &=& 0 \quad ; \quad \zeta^\bullet \,=\, 0
\end{eqnarray}
where $W(z)$ is a holomorphic superpotential whose norm (\ref{sectionnorm}) is assumed to be invariant under the isometries of the K\"ahler manifold that are gauged. The complex field dependent matrix $\mathcal{N}_{\Lambda\Sigma}(z,\bar{z})$ is an extra item of the construction that will appear in the lagrangian and it is supposed to transform  under an isometry of the K\"ahler manifold $z^i \mapsto f^i(z)$ according to the following fractional linear transformation:
\begin{equation}\label{gomorroida}
    \mathcal{N}(f(z),\bar{f}(\bar{z})) \, = \, \left( A_f \mathcal{N}(z,\bar{z})\, +\, B_f\right)\, \left( C_f \, \mathcal{N}(z,\bar{z}) \, + \, D_f\right)^{-1}
\end{equation}
the $2\, m \, \times \, 2\, m$ matrix:
\begin{equation}\label{cuccolino}
  \left(  \begin{array}{c|c}
       A_f & B_f \\
       \hline
       C_f & D_f
     \end{array}\right )
\end{equation}
being the image in a symplectic $2\,m$-dimensional linear representation of the element $f\in \mathrm{G}_{gauge}$ that generates the isometry  $z^i \mapsto f^i(z)$ of the K\"ahler manifold.
The numerical parameter $e$, corresponding to the overall scale of the superpotential (it can be reabsorbed by redefining $W^\prime \, = \, e \, W$) is a sort of gauge coupling constant of the $F$-gauging. Setting $e=0$ switches off the $S$ and $\mathcal{H}^i$ auxiliary fields (which present below) and their contribution to the scalar potential. In a way $e=0$ is just the meaning of New Minimal Formulation.
\par
Finally we should mention that the explicit values of the coefficients $\mu$ and $\nu$ were not calculated in \cite{castadauriafre2}. Although their determination is straightforward it is the result of a lengthy calculation that we skipped since the fermion bilinear contributions to the auxiliary fields are irrelevant for the goals of the present paper.
\par
The matter and gauge auxiliary fields have the following form:
\begin{eqnarray}
  \mathcal{H}^i &=& 2 \, e \,  \exp \, \left [ \ft 12 \, G\right ]\, g^{ij^\star} \, \partial_{j^\star} G \nonumber\\
  &=&  2 \, e \,  \exp \, \left [ \ft 12 \, W\right ]\, g^{ij^\star} \, \mathcal{D}_{j^\star} \overline{W}\label{Hauxfildo} \\
  D^\Lambda &=& \, g \,  \left(\mathrm{Im}\mathcal{N}^{-1}\right)^{\Lambda\Sigma}\, \mathcal{P}_\Sigma \label{Dfildo}
\end{eqnarray}
where $W$ is the already mentioned properly transforming superpotential and $\mathcal{P}_\Lambda$ are the momentum maps of the Killing vectors associated with the  gauged isometries of the K\"ahler manifold. The parameter $g$ is the gauge coupling constant, which is the $N=1$ theory is completely independent from $e$.
\subsection{The complete rheonomic Lagrangian}
Having twisted together the matter and the gauge multiplets in a complete Bianchi-consistent rheonomic parameterization of all the curvatures, the next step in the menu is the construction of a differential form written lagrangian whose superspace--extended field equations should reproduce such rheonomic parameterization and give also  the appropriate x-space equations of motion. This task was completed in \cite{castadauriafre2} up to the calculation of a few coefficients of four fermion terms, whose knowledge is superfluous for the sake of our present argument.
\par
Let us present such a Lagrangian. It can be regarded as the sum of three contribution corresponding respectively to the graviton sector, the WZ sector and the gauge sector:
\begin{equation}\label{lagrangianus}
    \mathcal{L} \, = \, \mathcal{L}_{Sugra} \, + \, \mathcal{L}_{WZ} \, + \, \mathcal{L}_{YM}
\end{equation}
Explicitly for the Sugra part we have:
\begin{equation}\label{lagsugra}
    \mathcal{L}_{Sugra} \, = \, \epsilon_{abcd}\, \mathfrak{R}^{ab}\, \wedge \, V^c \, \wedge \, V^d \, - \, 4 \, \left (\bar{\psi}^\bullet \, \wedge \, \gamma_a \,\rho_\bullet \, + \, \rho^\bullet \, \wedge \, \gamma_a \, \psi_\bullet \, \right ) \, \wedge \, V^a
\end{equation}
while the Lagrangian of the Wess Zumino multiplets is the following one:
\begin{eqnarray}
   \mathcal{L}_{WZ}&=& \, + \, \ft 23 \, g_{ij^\star} \, \left[Z^i_a \left(\nabla\,z^{j^\star} \, - \, \bar{\chi}^{j^\star}\, \psi^\bullet\right) \, + \, \bar{Z}^{j^\star}_a \left(\nabla\,z^{i} \, - \, \bar{\chi}^{i}\, \psi_\bullet\right)  \right]\,\wedge \, V_b \, \wedge \, V_c \, \wedge \, V_c \, \epsilon^{abcd}\nonumber \\
    && \, - \, \ft 16 \,  g_{ij^\star} \, Z^i_a \, \bar{Z}^{j^\star}_a \, \epsilon_{b_1\dots b_4} \, V^{b_1} \,\wedge \dots \wedge V^{b_4}\nonumber\\
  && - \, {\rm i} \, 2 \, g_{ij^\star} \, \left(\nabla \, z^i \, \wedge \, \bar{\chi}^{j^\star}\, \gamma_{ab} \, \psi^\bullet \, - \,
  \nabla \, z^{j^\star} \, \wedge \, \bar{\chi}^{i}\, \gamma_{ab} \, \psi_\bullet  \right) \, \wedge \, V^a \, \wedge \, V^b \nonumber\\
 && - \, {\rm i}\, 2 \,  g_{ij^\star} \, \bar{\chi}^i \, \gamma_a \chi^{j^\star} \, V^a \, \wedge \, \bar{\psi}_\bullet \, \wedge \, \gamma_b\, \psi^\bullet \, \wedge \, V^b \nonumber\\
  &&\, - \, \mathfrak{R}^a \, \wedge \, V_a \, \wedge  \, g_{ij^\star} \, \bar{\chi}^i \, \gamma_b \chi^{j^\star} \, V^b \nonumber\\
  && \, - \, \left[ \ft {1}{48}\, \left(g_{ij^\star} \, g_{k\ell^\star}\, + \, R_{j^\star i \ell^\star k}\right) \, \bar{\chi}^i \, \gamma^a \chi^{j^\star} \bar{\chi}^k \, \gamma_a \chi^{\ell^\star} \,\right.\nonumber\\
   &&\left.\, + \, V_{WZ}\, - \,\mathfrak{m}_{ij} \, \bar{\chi}^i \,\chi^j  \, - \, \mathfrak{m}_{i^\star j^\star}
   \, \bar{\chi}^{i^\star}\,\chi^{j^\star} \right ] \, \epsilon_{b_1\dots b_4} \, V^{b_1} \,\wedge \dots \wedge V^{b_4}\nonumber \\
  && \, - \, 4 \, \left( S \, \bar{\psi}^\bullet \, \wedge \, \gamma_{ab} \, \psi^\bullet \, + \,
  S^\star \, \bar{\psi}_\bullet \, \wedge \, \gamma_{ab} \, \psi_\bullet\right) \, \wedge \, V^a \, \wedge \, V^b \nonumber\\
 && \, + \, \, g_{ij^\star} \, \left(\mathcal{H}^i \, \bar{\chi}^{j^\star} \, \gamma^a \, \psi_\bullet \, + \,\mathcal{H}^{j^\star} \, \bar{\chi}^{i} \, \gamma^a \, \psi^\bullet \right) \, \wedge \, V^b \, \wedge \, V^c \, \wedge \,  V^d \, \epsilon_{abcd} \label{WZlagra}
\end{eqnarray}
Finally the Lagrangian for the gauge sector is the one below:
\begin{eqnarray}
 \mathcal{L}_{YM} &=& \, - \, \left ( \mathcal{N}_{\Lambda\Sigma}\, F^{+|\Lambda}_{ab}\, + \, \overline{\mathcal{N}}_{\Lambda\Sigma}\, F^{-|\Lambda}_{ab} \right) \, \left[\mathfrak{F}^\Sigma \, - \, {\rm i} \, \left(\lambda^\Sigma_\bullet \, \gamma_c \,\psi^\bullet \, + \,  \lambda^{\Sigma|\bullet} \, \gamma_c \,\psi_\bullet\right)\,\wedge \, V^c \,\right] \, \wedge \, V^a \, \wedge \, V^b \nonumber \\
 && \, - \, {\rm i} \, \ft 16 \, \left( \overline{\mathcal{N}}_{\Lambda\Sigma}\, F^{-|\Lambda}_{ab} \, F^{-|\Sigma}_{ab}\, - \,
  {\mathcal{N}}_{\Lambda\Sigma}\, F^{+|\Lambda}_{ab} \, F^{+|\Sigma}_{ab}\right) \, \epsilon_{c_1 \dots c_4} \, V^{c_1} \, \wedge \,  \dots \, \wedge \, V^{c_4}\nonumber\\
   && \, - \, {\rm i} \, \ft 13 \, \left( \mathcal{N}_{\Lambda\Sigma}\, \bar{\lambda}^\Lambda_\bullet \, \gamma^a \, \nabla\lambda^{\Sigma|\bullet} \, - \, \overline{\mathcal{N}}_{\Lambda\Sigma}\, \bar{\lambda}^{\Lambda|\bullet} \, \gamma^a \, \nabla\lambda^{\Sigma}_\bullet\right) \, \wedge \, V^b \, \wedge \, V^c \, \wedge \,  V^d \, \epsilon_{abcd} \nonumber\\
  && + {\rm i} \, q_1 \, \left( \mathcal{N}_{\Lambda\Sigma}\, \bar{\lambda}^\Lambda_\bullet \, \gamma^a \, \psi^\bullet \, D^\Sigma +
  \, \overline{\mathcal{N}}_{\Lambda\Sigma}\, \bar{\lambda}^{\Lambda|\bullet} \, \gamma^a \, \psi_\bullet \, D^\Sigma \right) \, \wedge \, V^b \, \wedge \, V^c \, \wedge \,  V^d \, \epsilon_{abcd} \nonumber\\
   && - \,{\rm i} \, 4 \, q_2 \, \mathfrak{F}^\Lambda \, \wedge \, \left(\overline{\mathcal{N}}_{\Lambda\Sigma}\, \bar{\lambda}^{\Sigma|\bullet}\, \gamma_a \, \psi_\bullet \, + \,  \mathcal{N}_{\Lambda\Sigma}\, \bar{\lambda}^{\Sigma}_{\bullet}\, \gamma_a \, \psi^\bullet\right) \, \wedge \, V^a \nonumber\\
  && \, - \, q_3 \, \mathfrak{R}^a \, \wedge \, V_a \, \wedge  \, \mathrm{Im}\mathcal{N}_{\Lambda\Sigma} \, \bar{\lambda}^\Lambda_\bullet \, \gamma_b \lambda^{\Sigma|\bullet} \, V^b \nonumber\\
 && \, - \, q_4 \, \bar{\psi}_\bullet \, \gamma^a \, \psi^\bullet \, \wedge \, V_a \, \wedge  \, \mathrm{Im}\mathcal{N}_{\Lambda\Sigma} \, \bar{\lambda}^\Lambda_\bullet \, \gamma_b \lambda^{\Sigma|\bullet} \, V^b \nonumber\\
   &&- \left[  \, V_{YM}\, - \,\mathfrak{m}_{i\Lambda} \, \bar{\chi}^i \,\lambda^\Lambda_\bullet  \, - \, \mathfrak{m}_{j^\star \Lambda} \bar{\chi}^{j^\star} \,\lambda^{\Lambda|\bullet} \, - \, \mathfrak{m}_{ \Lambda\Sigma} \,\bar{\lambda}^{\Lambda|\bullet}  \, \bar{\lambda}^{\Sigma|\bullet} \right.\nonumber\\
    &&\left.\, - \, \overline{\mathfrak{m}}_{ \Lambda\Sigma} \,\bar{\lambda}^{\Lambda}_{\bullet}  \, \bar{\lambda}^{\Sigma}_{\bullet}\right]  \,  \epsilon_{c_1 \dots c_4} \, V^{c_1} \, \wedge \,  \dots \, \wedge \, V^{c_4}\label{lagYM}
\end{eqnarray}
The coefficients $q_{1,2,3,4}$ are the undetermined coefficients mentioned above, while $\mathfrak{m}_{ij}$,
$\mathfrak{m}_{i^\star j^\star}$, $\mathfrak{m}_{i\Lambda}$, $\mathfrak{m}_{i^\star \Lambda}$ and $\mathfrak{m}_{\Lambda\Sigma}$ denote the mass-matrices of the fermions. Those that mix only the chiralinos were calculated in \cite{castadauriafre2} and have the following precise structure:
\begin{eqnarray}
  \mathfrak{m}_{ij} &=& \frac{e}{6}\,(\partial_{i}\,G\,\partial_{j}\,G \,+\,\nabla_i\,\partial_{j}\,G)\,e^{\frac{G}{2}}\\
  \mathfrak{m}_{i^\star j^\star} &=& \frac{e}{6}\,(\partial_{i^\star}\,G\,\partial_{j^\star}\,G \,+\,\nabla_{i^\star}\,\partial_{j^\star}\,G)\,e^{\frac{G}{2}}
\end{eqnarray}
for those that mix the chiralinos with the gauginos we can only mention the structure, but not the precise coefficients $q_{5,6,7}$ since they were not calculated in \cite{castadauriafre2}:
\begin{eqnarray}\label{gonzagariso}
    \mathfrak{m}_{i^\star \Lambda} & = & q_5 \, g^{ij^\star}\partial_{j^\star} \, G \, \mathcal{P}_\Lambda +q_6 \partial_i \mathcal{P}_\Lambda \nonumber\\
    \mathfrak{m}_{\Lambda\Sigma} & = & q_7 \exp\left[\ft 12 \, \mathcal{K}\right] \, \mathcal{P}_\Lambda \, \mathcal{P}_\Sigma
\end{eqnarray}
The  values of the coefficients $q_{5,6,7}$ are anyhow inessential for our present argument.
\subsection{The scalar potential}
The most important item for what we desire to discuss is the scalar potential. According to a general argument firstly introduced  in \cite{dewitNicolai} for the $\mathcal{N}=8$ theory and then extended to all the other theories  \cite{83cito,85cito,86cito}\footnote{For a general discussion see \cite{parlectures}.} the scalar potential in all supergravity theories can be represented as a quadratic form in the auxiliary fields appearing in the transformation rules of the spin $\ft 12$ and spin $\ft 32$ fermions with coefficients that are determined by the coefficients of the kinetic terms of the corresponding fermions. In the case of $\mathcal{N}=1$ supergravity this representation takes the following explicit form:
\begin{eqnarray}
  V &=& \underbrace{\ft 14 \, g_{ij^\star} \, \mathcal{H}^i \,\mathcal{H}^{j^\star}}_{\mbox{chiralinos contr.}} \, - \, \underbrace{3 \, S\, S^\star}_{\mbox{gravitino contr.}} \, + \,  \underbrace{\ft 13 \, \left(\mbox{Im} \,\mathcal{N}^{-1}\right)_{\Lambda\Sigma}\, D^\Lambda \, D^\Sigma }_{\mbox{gaugino contr.}} \nonumber \\
   &=& e^2 \exp\left[\mathcal{K}\right]\left(g^{ij^\star} \, \mathcal{D}_i\,W \, D_{j^\star} \overline{W}\, -\, 3\left| W \right|^2 \right) \, +\,
   \, \ft 13 \, g^2 \, \mbox{Im} \,\mathcal{N}_{\Lambda\Sigma}\, \mathcal{P}^\Lambda \, \mathcal{P}^\Sigma \label{masterformulone}
\end{eqnarray}
In this way we have reconstructed, within the  rheonomic framework, the general form of matter coupled $\mathcal{N}=1$ supergravity as first presented in \cite{standardN1}. The difference between the present formulation and that of \cite{standardN1} is mainly aesthetical. The  structures of the Hodge-K\"ahler geometry underlying the lagrangian had not yet been recognized at that time and many different items entering the lagrangian were  written in non covariant notations. The rheonomic approach, besides incorporating such geometrical structures in a natural way offers the unique possibility of obtaining the lagrangian directly in component form while being able at the same time to discuss such issues as the Old Minimal or New Minimal form of the supersymmetry algebra that, as we have advocated plays no real role in the embedding of cosmological models.
\par
The relevant point for us is that from standard $\mathcal{N}=1$ supergravity we can extract a bosonic lagrangian of the  form presented in the next section.
\subsection{General form of the bosonic lagrangian}
Relying on the discussions of the previous sections the general form of the bosonic lagrangian for an $\mathcal{N}=1$ supergravity is the following one:
\begin{eqnarray}
\mathcal{L}^{(4)} &=& \sqrt{|\mbox{det}\, g|}\left[{R[g]} - \frac{1}{2}
\nabla_{ {\mu}}z^{i}\, \nabla_{\nu}z^{j^\star} g_{ij^\star} \, g^{\mu\nu}
+ \,
\mbox{Im}\mathcal{N}_{\Lambda\Sigma}\, F_{ {\mu} {\nu}}^\Lambda
F^{\Sigma| {\mu} {\nu}} \, - \, V\right]
\nonumber\\
&&
+
\frac{1}{2}\mbox{Re}\mathcal{N}_{\Lambda\Sigma}\, F_{ {\mu} {\nu}}^\Lambda
F^{\Sigma}_{ {\rho} {\sigma}}\epsilon^{ {\mu} {\nu} {\rho} {\sigma}}\,,\nonumber\\
\label{d4generlag}
\end{eqnarray}
where
\begin{equation}\label{framartinocampanaro}
    \nabla_{ {\mu}}z^{i} \, = \, \partial_\mu z^i \, + \, g \, \mathcal{A}_\mu^\Lambda \, k^i_\Lambda(z)
\end{equation}
and the potential is given by eq.(\ref{masterformulone}).
\par
This is our starting point in order to revisit the ideas introduced in \cite{minimalsergioKLP} showing that by means of a translation gauging we can embed into standard supergravity all the integrable cosmologies discovered and classified in \cite{noicosmoitegr}.
\section{Embedding inflaton models}
Starting from a generic theory of supergravity with $n+1$ Wess-Zumino multiplets and $m$ vector multiplets we focus on the case where the K\"ahler manifold has the following direct product structure:
\begin{equation}\label{direttoprodo}
    \mathcal{M}_{\mbox{K\"ahler}} \, = \, \mathcal{M}_{J} \, \otimes \, \mathcal{M}_{\mathcal{K}}
\end{equation}
the submanifold $\mathcal{M}_{J} $ having complex dimension one while the manifold $\mathcal{M}_{\mathcal{K}}$ has complex dimension $n$. Furthermore denoting by $\Omega$ the complex coordinate that labels the point of $\mathcal{M}_{J} $ and writing:
\begin{equation}\label{omegaCB}
    \Omega \, = \, {\rm i} C \, + \, B
\end{equation}
we assume that the group of translations $B\rightarrow B+c$ is a group of isometries for the K\"ahler manifold $\mathcal{M}_{J}$. All this  amounts to assume the following structure for the complete K\"ahler potential of the manifold (\ref{direttoprodo}).
\begin{equation}\label{fustacchio}
   \widehat{\mathcal{K}} \, = \, J \left(\mbox{Im}\Omega\right) \, + \, \mathcal{K}(z,z)
\end{equation}
where $J \left(\mbox{Im}\Omega\right)$ is any real  function of its argument while $\mathcal{K}(z,z)$  is a generic K\"ahler potential for the manifold $\mathcal{M}_{\mathcal{K}}$.
\par
The K\"ahler metric that defines the kinetic terms of the scalars in the lagrangian takes the form:
\begin{equation}\label{menopausa}
    ds^2_{\mbox{K\"ahler}} \, = \, \ft 14 \, J^{\prime\prime}\left(\mbox{Im}\Omega\right) \left| d\Omega\right|^2 \, + \, g_{ij^\star}\, dz^i \, dz^{j^\star}
\end{equation}
and the $G$-function is the following one:
\begin{equation}\label{ficino}
    G \, = \, J\left(\mbox{Im}\Omega\right) \, + \, \mathcal{K}(z,z) \, + \, \log \left[ W(\Omega,z) \, \bar{W}(\bar{\Omega},\bar{z}) \right]
\end{equation}
The mechanism that allows to generate an inflaton potential is based on the gauging of the translation isometry $\Omega \to \Omega + c$ with $c\in \mathbb{R}$. In order for this gauging to be feasible it is necessary that such an isometry of the K\"ahler metric extends to a \textit{bona fide} global symmetry of the entire supergravity lagrangian coupled to the $n+1$ Wess-Zumino multiplets. This happens if and only if the function $G$ is invariant under the translation $\Omega \to \Omega + c$. Such invariance implies that the superpotential $W(\Omega,z)$ should not depend on the field $\Omega$:
\begin{equation}\label{decuppiu}
    \partial_\Omega \, W(\Omega,z) \, = \, 0 \quad \Rightarrow \quad W(\Omega,z) \, = \, W(z)
\end{equation}
Under these conditions the Wess-Zumino part of the complete scalar potential takes the following form:
\begin{equation}\label{gozzo}
    V_{WZ} \, = \, \exp\left[J\right] \,
    \left[\underbrace{\exp\left[\mathcal{K}\right]\left(g^{ij^\star} \, \mathcal{D}_i\,W \, D_{j^\star} \overline{W}\, -\, 3\left| W \right|^2 \right)}_{\mbox{potential of the $n$ multiplets}}\, +\,
    \exp\left[ \mathcal{K}\right] \frac{(J^\prime)^2}{J^{\prime\prime} } \, \left|W\right|^2 \right]
\end{equation}
In addition, the gauging of the translation symmetry produces a contribution to the full potential:
\begin{equation}\label{fullopotto}
    V \, = \, V_{YM} \, + \, V_{WZ}
\end{equation}
that according to the general formula (\ref{masterformulone}) has the following structure:
\begin{equation}\label{additional}
    V_{YM} \, = \, \ft 13 \, \mbox{Im} \, \mathcal{N}_{00} \, \left( \mathcal{P}^0 \right)^2
\end{equation}
the function $\mathcal{P}^0$ being the momentum map of the Killing vector that generates the translation isometry $\Omega \to \Omega +c$. In the case in which we choose the certainly invariant function:
\begin{equation}\label{corifeo}
    \mathcal{N}_{00} \, = \, \delta_{\Lambda\Sigma}
\end{equation}
the gauge contribution to the potential becomes the perfect square of a function of the coordinate $C$ which is very simple to calculate. Using the general formula (\ref{bozhemoi}) we find:
\begin{eqnarray}\label{fanciulla}
    \mathcal{P}_0 & = & \ft 12 \, \frac{\mathrm{d}J}{\mathrm{d}C} \, \equiv \,  \ft 12 \, J^\prime \nonumber\\
    V_{YM} & = & \ft {1}{12} \, g^2 \, \left(J^\prime(C) \right)^2
\end{eqnarray}
The complete potential (\ref{fullopotto}) can be reduced to a function of the single field $C$ if the other moduli fields
$z^i$ can be stabilized in a $C$-independent way. A very simple calculation, shows that this occurs generically if the superpotential $W(z)$ has a critical point at $W=0$, namely if a set of values $z_0^i$ does exist such that:
\begin{equation}
 \left.   \mathcal{D}_i \,W \right|_{z^i=z^i_0} \, = \, 0 \quad ; \quad W \left (  z_0 \right ) \, = \, 0 \label{modstab}
\end{equation}
After stabilization the potential reduces to:
\begin{equation}\label{VdiC}
    V(C) \, = \, \mbox{const}^2 \, \times  \, \left( J^\prime(C) \right)^2
\end{equation}
which is definite positive. The kinetic term, however, is not canonical, since, after neglecting the $B$ field that is eaten up by the vector field
in the Higgs mechanism, the relevant scalar part of the Lagrangian reduces to:
\begin{equation}\label{guterlat}
    \mathcal{L}_{scalar} \, = \, 2 \, \left ( \ft 14  \, J^{\prime\prime} (C) \, \partial^\mu C \, \partial_\mu C \, - \, \mbox{const}^2 \, \times  \, \left( J^\prime(C) \right)^2\right)
\end{equation}
Following the basic suggestion of \cite{minimalsergioKLP} we can introduce a field redefinition that reduces the kinetic term to its canonical form. Setting
\begin{equation}\label{balucco}
    C \, = \, C(\phi)
\end{equation}
imposing that:
\begin{equation}\label{cundiziunu}
    J^{\prime\prime} (C) \, (\mathrm{d}C)^2 \, = \, J^{\prime\prime} (C) \, \left(\frac{\mathrm{d}C}{\mathrm{d}\phi}\right)^2\, (\mathrm{d}\phi)^2\, = \, (\mathrm{d}\phi)^2
\end{equation}
Recalling equation (\ref{fanciulla}) and naming:
\begin{equation}\label{fratellodiborsa}
    D(\phi) \, = \, - \, 2 \, \mathcal{P}_0(C) \, = \, J^\prime(C)
\end{equation}
condition (\ref{cundiziunu}) becomes:
\begin{equation}\label{figliodicane}
1 \, = \,    \frac{\mathrm{d}D(\phi)}{\mathrm{d}\phi} \, \frac{\mathrm{d}\phi}{\mathrm{d}C} \left(\frac{\mathrm{d}C}{\mathrm{d}\phi}\right)^2 \, = \, \frac{\mathrm{d}D(\phi)}{\mathrm{d}\phi} \, \left(\frac{\mathrm{d}C}{\mathrm{d}\phi}\right)
\end{equation}
from which we deduce
\begin{equation}\label{rosalba}
    \left(\frac{\mathrm{d}C}{\mathrm{d}\phi}\right) \, = \, \left( \frac{\mathrm{d}D(\phi)}{\mathrm{d}\phi}\right)^{-1}
\end{equation}
In this way we obtain:
\begin{equation}\label{gagliardo}
  \mathcal{L}_{scalar} \, = \, 2 \, \left ( \ft 14  \,  \partial^\mu \phi \, \partial_\mu \phi \, - \, \mbox{const}^2 \, \times  \, \left( D(\phi) \right)^2\right)
\end{equation}
In this way, any \textit{ positive definite potential }$V(\phi)$ can be embedded into supergravity by reinterpreting its square root as the momentum map of a translation killing vector in a one-dimensional K\"ahler manifold \footnote{For simplicity we have dropped the index $0$ since there is only one gauged generator and only one prepotential, no confusion being possible.}:
\begin{equation}\label{giocone}
    D(\phi) \, = \, \sqrt{V(\phi)} \quad \rightarrow \quad \mathcal{P}_0 \, \equiv \, \mathcal{P}\, =\,- \, \ft 12 \, \sqrt{V(\phi)}
\end{equation}
The intriguing question is how to work out the corresponding K\"ahler potential or, formulating it in a more intrinsic way, independently from the used coordinates, \textit{what is the geometry of the underlying one-dimensional K\"ahler manifold?}
\subsection{Relation between the potential and the K\"ahler curvature}
In order to give an answer to the above question, let us recall that the local geometry of a two dimensional manifold with euclidian signature is completely determined by a single invariant function of the two coordinates, namely by the unique intrinsic component of the  Riemann tensor. Notwithstanding the fact that the calculation of the K\"ahler potential is quite difficult for a generic potential $V(\phi)$ since it involves the use of the inverse of functions defined by indefinite integrals, yet the calculation of the intrinsic curvature is fairly easy and leads to a very suggestive relation which can shed light on the geometrical interpretation of inflaton models. Let us see how this is possible.
\par
Reinstalling the axion $B$ and using the relations (\ref{figliodicane}) and (\ref{rosalba}) we see that, in terms of the coordinates $\{\phi,B\}$ the metric of the K\"ahler manifold has the following form:
\begin{equation}\label{metricozza}
    ds^2_{\mbox{K\"ahler}} \, = \, \ft 14 \,\mathrm{d}\phi^2 \, + \, \left(\mathcal{P}^\prime(\phi)\right)^2 \, dB^2
\end{equation}
where the momentum map of the translation Killing vector $\mathcal{P}^\prime(\phi)$  is  related to the potential by eq.(\ref{giocone}).
This coordinate is well adapted to translation symmetry $B\to B+ c$ which is manifest. We introduce the zweibein $E^0 \, E^1$ by setting:
\begin{eqnarray}
  E^1 &=& \ft 12 \, \mathrm{d}\phi \nonumber \\
  E^2 &=&  \mathcal{P}^\prime(\phi) \, \mathrm{d}B \label{zweibein}
\end{eqnarray}
and the Levi-Civita spin connection trough the standard structural equations:
\begin{eqnarray}
  \mathrm{d}E^1 \, + \, \omega \, \wedge \, E^2&=& 0 \\
  \mathrm{d}E^2 \, - \, \omega \, \wedge \, E^1 &=& 0 \label{levicivspincon}
\end{eqnarray}
Eq.s (\ref{levicivspincon}) are immediately solved by
\begin{equation}\label{cristubulu}
    \omega \, = \, -2 \, \mathcal{P}^{\prime\prime}(\phi) \, \mathrm{d}B
\end{equation}
From this result immediately follows the curvature two form:
\begin{eqnarray}
    \mathfrak{R} & \equiv & \mathrm{d}\omega \, \equiv \, R(\phi) \, E^1 \, \wedge \, E^2 \nonumber\\
    R(\phi) & = & - 4 \, \frac{\mathcal{P}^{\prime\prime\prime}(\phi)}{\mathcal{P}^{\prime}(\phi)}\label{garducci}
\end{eqnarray}
In terms of the scalar potential $V(\phi)$ we obtain the following relation:
\begin{equation}\label{giunone}
     R(\phi) \, = \, -\, 4 \, \left ( \frac{V^{\prime\prime\prime}}{V^{\prime}} \, - \, \ft 32 \, \frac{V^{\prime\prime}}{V} \, - \, \ft  34 \, \left( \frac{V^\prime}{V}\right)^2 \right)(\phi)
\end{equation}
The underlying K\"ahler manifold is a homogeneous space $\frac{\mathrm{SU(1,1)}}{\mathrm{U(1)}}$ only when the curvature $R(\phi)$ turns out to be a constant. This can happen also with non trivial potentials $V(\phi)$ which therefore simply emerge from an astute change of coordinates. When $R(\phi)$ is not constant the manifold is necessarily non homogeneous and one can try to study its properties from the behavior of its curvature applying case by case standard techniques of differential geometry. Understanding the intrinsic geometric properties of the K\"ahler manifold associated with each considered inflaton potential is the preliminary and main step in order to understand its possible dynamical origin in string theory.
\par
As a final general remark let us mention that starting from the form (\ref{metricozza}) of the K\"ahler metric we can easily derive a parametric expression for the geodesics reduced to quadratures.  The analysis of the shape of geodesics provides a visualization of the geometry of a space and the following formulae might be very useful to investigating the elusive nature of the new spaces associated  with the inflationary potential.
\par
Addressing the calculation of  geodesics as a variational problem where the metric plays the role of a Lagrangian and using as affine parameter the geodesic length $s$ we write:
\begin{equation}\label{lagrageo}
    \mathcal{L} \, = \, \dot{\phi}^2\, + \, Q(\phi)^2 \dot{B}^2
\end{equation}
where we have set $\dot{\phi} \, = \, \frac{\mathrm{d}\phi}{\mathrm{d}s}$ and $\dot{B} \, = \, \frac{\mathrm{d}B}{\mathrm{d}s}$. Furthermore we have  denoted:
\begin{equation}\label{gillo}
    Q(\phi) \, = \, \frac{\mathrm{d}}{\mathrm{d}\phi} \, \sqrt{V(\phi)}
\end{equation}
The variable $B$ is cyclic and we have an immediate integral of motion that we name $1/R$:
\begin{equation}\label{sicilia}
    Q(\phi)^2 \, \dot{B} \, = \, \frac{1}{R}
\end{equation}
Replacing this information back in the Lagrangian we have the consistency constraint
\begin{equation}\label{costretto}
    1 \, = \, \dot{\phi}^2\, + \, \frac{1}{R^2} \, \frac{1}{Q^2(\phi)}
\end{equation}
Using the relation $ \frac{\mathrm{d}\phi}{\mathrm{d} s} \, = \dot{\phi} \, = \, \frac{\mathrm{d}\phi}{\mathrm{d} B} \, \frac{\mathrm{d}B}{\mathrm{d}s}$ and inserting once more the integral of motion (\ref{sicilia}) we obtain:
\begin{equation}\label{ficino}
    B(\phi) \,  = \, \pm\,\int \frac {d\phi}{Q(\phi) \, \sqrt{R^2 \, Q^2(\phi)\, - \, 1}}\,  + \, B_0 \,
\end{equation}
where $B_0$ is an integration constant. The function $B(\phi)$  is the real part of complex coordinate $\Omega$. By $C$ we have denoted the imaginary part of the same for which we already know its expression in terms of the parameter $\phi$. Indeed in  previous calculations we have found:
\begin{equation}\label{marsilio}
    C(\phi) \, = \, \int \frac {d\phi}{Q(\phi)}
\end{equation}
We have therefore a complete set of geodesics. They are labeled by the parameters $R$ and $B_0$ and are given in parametric form in terms of $\phi$. In the case of the Lobachevski plane $Q(\phi) \, = \, \exp[-\phi]$, and by straightforward calculation we obtain $C(\phi) \, = \, \exp[\phi]$ and $B(\phi) \, = \, \pm\,\sqrt{R^2 -C^2} +B_0$, so that the geodesics are half circles in the upper complex plane with radius $R$ and center on the real axes in $B_0$. With different functions $Q(\phi)$ we obtain that the geodesics are deformations of such circles.
\section{Examples with integrable potentials in supergravity}
In a recent paper \cite{noicosmoitegr} A. Sagnotti and the two of us have presented a bestiary  of one-field potentials that lead to integrable cosmological models, discussing also several properties of the ensuing exact solutions. It was pointed out that some of these models display phenomenologically attractive features, in some cases yielding a graceful exit from inflation. In a couple of separate publications  Sagnotti has also shown that the phenomenon of climbing scalars, displayed by all of the integrable models we were able to classify, has the potential ability to explain the  oscillations in the low angular momentum part of the CMB  spectrum, apparently observed by PLANCK. In his recent talk given at the Dubna SQS2013 workshop, our coauthor has also shown a best fit to the PLANCK data for the low $\ell$ part of the spectrum, by using the series of integrable potentials\footnote{In comparing the following equation with the table of paper \cite{noicosmoitegr}, please note the coefficient $\sqrt{3}$ appearing in the exponents that has been introduced to convert the unconventional normalization of the field $\varphi$ used there to the canonical normalization of the field $\phi$ used here.}
\begin{equation}\label{gammaserie}
    V(\phi) \, = \, a \, \exp\left[ 2\, \sqrt{3} \, \gamma \, \phi\right] + b \, \exp\left[  \sqrt{3} \, (\gamma +1)\, \phi\right]
\end{equation}
This best fit selects the particularly nice value $\gamma \, = \, -\ft 76$.
\par
In paper \cite{noicosmoitegr} we posed the question whether integrable potentials can be fitted into supergravity and in a forthcoming publication \cite{nointegrable2} we show that, although their type is very natural in gauged extended supergravities, the precise combinations implied by integrability are hard to be met. For instance, by means of the classification of all the gaugings of the  $STU$ model with $\mathcal{N}=2$ supersymmetry, in  \cite{nointegrable2} we are able to exclude the presence of integrable potentials in such an  environment. However, although hard, the task is not impossible and, within $\mathcal{N}=1$ supergravity, gauged by superpotentials of the type that appear in flux compactifications, we were able to single out a pair of integrable cases (they will be presented in \cite{nointegrable2}).
\par
On the other hand, adopting the point of view presented in this paper, which springs from the ideas put forward in \cite{minimalsergioKLP}, the embedding of integrable potentials into supergravity becomes feasible for all the cases where the potential is positive definite. This is certainly the case for the best fit case of eq.(\ref{gammaserie}) when $a >0,b>0$ and for almost all cases in our bestiary.
\par
In this framework, the effort to understand the Physics underlying the emergence of such integrable potentials changes gear. Instead of looking for the mechanisms that determine suitable superpotentials, the focus is shifted on trying to understand the nature of the corresponding one dimensional K\"ahler geometry. In the road toward the solution of this problem the formula (\ref{giunone}) that relates the potential to the curvature  of the K\"ahler manifold constitutes a first illuminating step. Evaluating it on a few examples of scalar potentials we see that the underlying K\"ahler manifold is asymptotically (for large positive or negative $\phi$) a coset manifold $\frac{SU(1,1)}{U(1)}$, however, typically with different values of its curvature at one and the other extremum of the range of $\phi$. In this way it appears that the manifold sustaining such potentials are a kind of instanton connecting two different vacua. All this reminds us of the phenomena taking place in the case of Calabi-Yau moduli space where the geometry is, for large radii, that of a homogeneous space and it is instead deformed at small radii by non perturbative quantum corrections induced by world-sheet instantons. The shift of the potential from the F-sector of the superpotential to the D-sector and the K\"ahler potential also vaguely reminds us of mirror symmetry and the trading of complex structure deformations with K\"ahler class deformations, alias of the exchange of the A-twisted and B-twisted topological field theories. Further inspiration from these vague analogies might help to get a more profound understanding of the integrable potentials at stake and of the corresponding K\"ahler geometries. In this paper we confine ourselves to mention such possibility and we just present a first glance at the curvatures corresponding to a few examples of integrable potentials.
\subsection{The integrable series 2 in \cite{noicosmoitegr}}
\begin{figure}[!hbt]
\begin{center}
\iffigs
 \includegraphics[height=60mm]{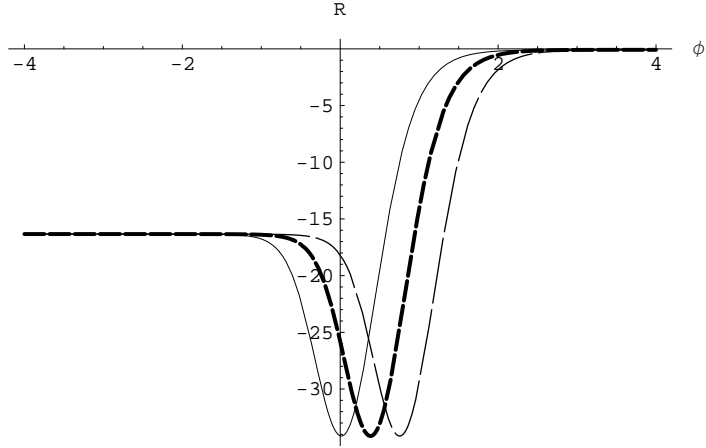}
\else
\end{center}
 \fi
\caption{\it
Plots of the K\"ahler curvature, versus coordinate $\phi$ for the best fit model $\gamma \, = \, -\ft 76$. The continuous line corresponds to the value $\lambda = \ft 14$, the thick dashed line corresponds to the value $\lambda = 1$, while the thin dashed line corresponds to the value $\lambda = 4$. The smaller is $\lambda$ the earlier occurs the peak in  the curvature. The bigger $\lambda$ the longer the
initial value of the curvature $R= \ft {49}{12}$ is maintained and the later we see the peak.}
\label{settesestiR}
 \iffigs
 \hskip 1cm \unitlength=1.1mm
 \end{center}
  \fi
\end{figure}

The integrable potentials of the series 2 in the bestiary of \cite{noicosmoitegr} are those mentioned in eq.(\ref{gammaserie}). Applying eq.(\ref{giunone}) to these potentials we find the following expression for the
K\"ahler curvature:
\begin{eqnarray}
  R_\gamma(\phi) &=& \, - \, 4 \, \frac{\mathrm{N}(\phi)}{\mathrm{D}(\phi)} \nonumber\\
  \mathrm{N}(\phi) &=& 3 \left(8 a^3 e^{6 \sqrt{3}
   \gamma  \phi } \gamma ^3+b^3
   e^{3 \sqrt{3} (\gamma +1)
   \phi } (\gamma +1)^3+4 a^2 b
   e^{\sqrt{3} (5 \gamma +1)
   \phi } \left(5 \gamma
   ^3+1\right)\right.\nonumber\\
   &&\left.+2 a b^2 e^{2
   \sqrt{3} (2 \gamma +1) \phi }
   \left(8 \gamma ^3-3 \gamma
   ^2+6 \gamma +1\right)\right)\nonumber\\
  \mathrm{D}(\phi)&=& 4 \left(e^{2 \sqrt{3} \gamma
   \phi } a+b e^{\sqrt{3}
   (\gamma +1) \phi }\right)^2
   \left(2 a e^{2 \sqrt{3}
   \gamma  \phi } \gamma +b
   e^{\sqrt{3} (\gamma +1) \phi
   } (\gamma +1)\right)\label{gammaserioso}
\end{eqnarray}
If  in eq.(\ref{gammaserioso}) we insert the best fit value $\gamma \, = \, - \ft 76$ found by Sagnotti and furthermore we redefine the parameters setting $a=\lambda*b$, we obtain:
\begin{eqnarray}
  R_{-\ft 76}(\phi) &=& \, - \, \frac{1}{12} \left(39 \lambda
   \left(\frac{280}{14 \lambda
   +e^{\frac{13 \phi }{2
   \sqrt{3}}}}-\frac{15 \lambda
   +28 e^{\frac{13 \phi }{2
   \sqrt{3}}}}{\left(\lambda
   +e^{\frac{13 \phi }{2
   \sqrt{3}}}\right)^2}\right)+1
   \right)
\end{eqnarray}
where the overall scale $b$ cancels. For all $\lambda$.s the function $R_{-\ft 76}(\phi) $ has the property:
\begin{equation}\label{giocattolo}
    R_{-\ft 76}(-\infty) \, = \, - \, \frac{49}{3} \quad ; \quad R_{-\ft 76}(\infty) \, = \, - \, \frac{1}{12}
\end{equation}
The structure of the curvature functions, whose plots are displayed for some values of $\lambda$ in fig.\ref{settesestiR}, reveals that these spaces realize a smooth transition from an $\mathrm{SU(1,1)}/\mathrm{U(1)}$ with one value of the curvature $R_{-\infty}$ to another one with a different value $R_{\infty}$. The shape of that transition has an interesting wiggle structure with first a peak and then a rapid descent to the lower value of the curvature.
Let us stress that in terms of the K\"ahler potential for $\mathrm{SU(1,1)}/\mathrm{U(1)}$ written as:
\begin{equation}\label{su11callo}
    \mathcal{K}_{\su(1,1)} (z,\bar{z}) \, = \, - \, \log \, \left [(z \, - \, \bar{z})^q \right]
\end{equation}
the curvature is simply linked to the parameter $q$ as it follows:
\begin{equation}\label{Rqrela}
    R_{\su(1,1)} \, =\, - \, \frac{4}{q}
\end{equation}
Indeed, using the solvable parameterization of the manifold
\begin{equation}\label{faccina}
    z\, = \, {\rm i} \, \exp\left[\frac{1}{\sqrt{q}} \,\phi\right] \, + \, B
\end{equation}
the $\mathrm{SU(1,1)}/\mathrm{U(1)}$ metric is written in the form (\ref{metricozza}) with:
\begin{equation}\label{marsilio}
    \mathcal{P}^\prime (\phi)\, = \, \sqrt{\frac{q}{4}} \, \exp\left [ - \,\frac{1}{\sqrt{q}} \,\phi\right]
\end{equation}
which upon use of eq.(\ref{garducci}) yields the result (\ref{Rqrela}). In conclusion the best fit K\"ahler manifold is a kink connecting the two $q$-indices:
\begin{equation}\label{figliomio}
    q_{-\infty} \, = \, \frac{12}{49} \quad \Rightarrow \quad q_{\infty} \, = \, 48
\end{equation}
These rational numbers are likely to hide some profound meaning in terms of brane-wrapping or similar higher dimensional mechanisms as suggested by the brane interpretation of the best fit model put forward by Sagnotti and briefly summarized in \cite{noicosmoitegr}.
\subsection{The integrable series 7 in \cite{noicosmoitegr}}
\begin{figure}[!hbt]
\begin{center}
\iffigs
 \includegraphics[height=60mm]{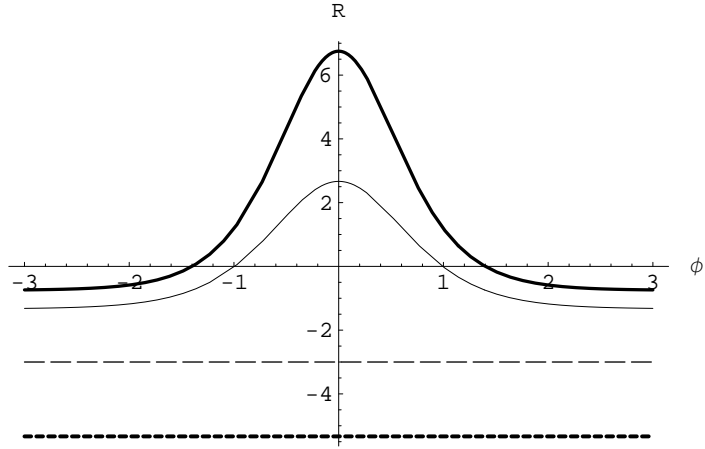}
\else
\end{center}
 \fi
\caption{\it
Plots of the K\"ahler curvature, versus coordinate $\phi$ for the integrable potential of eq. (\ref{goodwishes}). The continuous thin line corresponds to the value $\gamma = \ft 23$, the thick continuous  line corresponds to the value $\gamma = \ft 34$.  The thin dashed line corresponds to the value $\gamma = \ft 12$ while the thick dashed line corresponds to the value $\gamma = \ft 13$ which yields the very special integer value $q=3$.}
\label{coshgammus}
 \iffigs
 \hskip 1cm \unitlength=1.1mm
 \end{center}
  \fi
\end{figure}

The integrable potentials of the series 7 in the bestiary of \cite{noicosmoitegr} are quite interesting since when we put either $C_1 = 0$ or $C_2 =0$ they take the form of a perfect square, as suggested by their momentum map interpretation. Here we consider the case $C_2 \, = \, 0$ for exemplification. Using the appropriate conversion of normalizations we have:
\begin{eqnarray}
    V(\phi) &= & \mbox{const} \, \left( \mathcal{P}(\phi) \right)^2 \nonumber\\
    \mathcal{P}(\phi) & = & -\frac{1}{2} \cosh
   ^{\frac{1}{\gamma
   }-1}\left(\sqrt{3} \gamma
   \phi \right) \label{goodwishes}
\end{eqnarray}
Using this information in the curvature formula eq.(\ref{garducci}) we find:
\begin{equation}\label{fluorescenza}
    R_{\gamma|\cosh}(\phi) \, = \, - \, 12 \, \left(\left(6 \gamma ^2-5
   \gamma +1\right) \tanh
   ^2\left(\sqrt{3} \gamma  \phi
   \right)+(3-5 \gamma ) \gamma
   \right)
\end{equation}
In this case the asymptotic value of the curvature is the same at $\phi\, = \, \pm \infty$ and we have:
\begin{equation}\label{fortepiano}
    R_{\gamma|\cosh}(\pm \infty) \, = \, -12 \, (1\, -\, \gamma)^2 \quad \Rightarrow \quad q_{\pm\infty} \, = \, 3 \,(1\, -\, \gamma)^2
\end{equation}
A very intriguing fact is that for the two values $\gamma= \ft 12$ and $\gamma = \ft 13$ the dependenc on $\phi$ of the curvature disappears. This means that in these two cases the K\"ahler manifold is just the homogeneous space $\mathrm{SU(1,1)}/\mathrm{U(1)}$ and that the potential is created by gauging some appropriate translational subgroup conjugate to the standard one $\left(
                                                                          \begin{array}{cc}
                                                                            1 & c \\
                                                                            0 & 1 \\
                                                                          \end{array}
                                                                        \right)$ of $\mathrm{SL(2,\mathbb{R})}\sim \mathrm{SU(1,1)}$.
The plots of the K\"ahler curvature for some different values of $\gamma$ are displayed in fig.\ref{coshgammus}.
For all the  $\gamma$.s different from the two critical values the plots shows that also in this case two asymptotically homogeneous K\"ahler manifolds are smootly connected. The only difference with the previous case is that they have the same curvature at both end points of the range.
\subsection{The $\mathrm{ArcTan}$ potential 6 in \cite{noicosmoitegr}}
In the bestiary compiled in \cite{noicosmoitegr} a distinct place is occupied by the following potential 6:
\begin{equation}\label{arcotannus}
    V_{ArcTan}(\phi) \, = \, \arctan\left ( \exp\left[-\, 2\, \sqrt{3} \, \phi\right]\right)
\end{equation}
As shown in \cite{noicosmoitegr}, the exact solutions streaming from such a potential display an essentially realistic number $e$-fold inflation followed by a graceful exit to a power-like type of expansion. It is therefore quite challenging trying to understand which type of K\"ahler manifold allows its embedding into $\mathcal{N}=1$ supergravity via the discussed gauging mechanism. As a first step in this direction we present the curvature of such a K\"ahler manifold. It is given by the following function:
\begin{eqnarray}\label{goodpoint}
   R_{ArcTan}(\phi) & = & \, -\, 4\, \frac{\mathrm{N}(\phi)}{\mathrm{D}(\phi)} \nonumber\\
   \mathrm{N}(\phi) & = & 3 e^{4 \sqrt{3} \phi }
   \left(8 \cosh \left(4
   \sqrt{3} \phi \right) \tan
   ^{-1}\left(e^{-2 \sqrt{3}
   \phi }\right)^2-24 \tan
   ^{-1}\left(e^{-2 \sqrt{3}
   \phi }\right)^2\right.\nonumber\\
   &&\left.-12 \sinh
   \left(2 \sqrt{3} \phi \right)
   \tan ^{-1}\left(e^{-2
   \sqrt{3} \phi
   }\right)+3\right)\nonumber\\
  \mathrm{D}(\phi)& = & \left(1+e^
   {4 \sqrt{3} \phi }\right)^2
   \tan ^{-1}\left(e^{-2
   \sqrt{3} \phi }\right)^2
\end{eqnarray}
The limiting values of this curvature at $\phi \, = \, \pm \infty$ are the following quite inspiring integer ones:
\begin{eqnarray}
  R_{ArcTan}(-\infty) &=& \, - \, 48 \, \quad \Rightarrow \quad q_\infty \, = \, \ft 13\nonumber\\
  R_{ArcTan}(\infty) &=& \, - \, 12 \, \quad \Rightarrow \quad q_\infty \, = \, \ft 43\nonumber\\
\end{eqnarray}
\begin{figure}[!hbt]
\begin{center}
\iffigs
 \includegraphics[height=60mm]{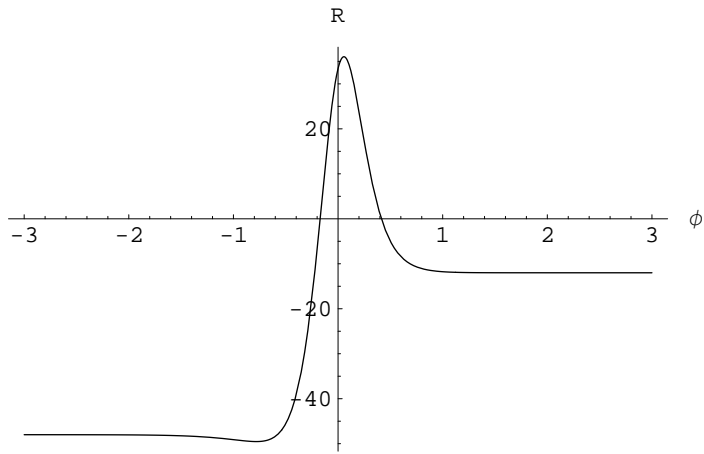}
\else
\end{center}
 \fi
\caption{\it
Plot of the K\"ahler curvature, versus the coordinate $\phi$ for the integrable potential of eq. (\ref{arcotannus}) . }
\label{arcotangente}
 \iffigs
 \hskip 1cm \unitlength=1.1mm
 \end{center}
  \fi
\end{figure}
The plot of the curvature, displayed in fig.\ref{arcotangente}, apart from the two asymptotic behaviors shows an interesting oscillation in the interior which certainly demands further consideration and interpretation.
\section{Conclusions}
In this paper we have shown that the Copernican revolution about the embedding into supergravity of inflationary potentials that was started by the authors of \cite{minimalsergioKLP} can be naturally formulated within Standard N=1 supergravity, the use of the New Minimal formulation and of conformal tensor calculus being unessential. We have also shown that each potential defines a new K\"ahler geometry whose main invariant classifier, the curvature, is accessible to calculation via a nice formula. Also the geodesics of the space defined by every potential are accessible and given in parametric form by quadratures. We advocate that the study of these geometries is the most urgent task in order to deepen our understanding of inflationary cosmologies.
\par
The most relevant consequence of the Copernican revolution is that by this token all the integrable potentials of the bestiary compiled in \cite{noicosmoitegr} become accessible to supergravity.
\par
We have shortly dwelled on the embedding and on the curvature of one of the integrable potentials which the analysis of Sagnotti
shows to be the most promising for the explanation of PLANCK data at low angular momenta $\ell$. The shape of the curvature for this model is very simple and inspiring.
We conclude by saying that a new unexpected but potentially very profound path to connect inflation to fundamental theory has opened up. The developments might be quite far reaching.

\section*{Acknowledgments}
The work of A.S. was supported in part by the RFBR Grants No. 11-02-01335-a, No. 13-02-91330-NNIO-a and No. 13-02-90602-Arm-a.
We express our gratitude to our friends and collaborators A. Sagnotti and M. Trigiante for very useful and inspiring comments.

\end{document}